\documentstyle[12pt,epsf]{article}
\newcommand{\be}{\begin{equation}}
\newcommand{\ee}{\end{equation}}
\newcommand{\ben}{\begin{eqnarray}}
\newcommand{\een}{\end{eqnarray}}
\def\simlt{\rlap{\lower 3.5 pt\hbox{$\mathchar \sim$}}\raise 1pt \hbox {$<$}}

\begin{document}
\title{
\vspace{-3.0cm}
\begin{flushright}
{\normalsize KEK-CP-063}\\
\vspace{-0.3cm}
{\normalsize October 1997 }\\
\end{flushright}
\vspace*{2.0cm}
{\Large Scaling Study of the Two-Flavor Chiral Phase Transition with the
Kogut-Susskind Quark Action in Lattice QCD}\vspace*{0.5cm}}

\author{\centerline{\Large JLQCD Collaboration}\\ \\
          S. Aoki$^{a}$,  M. Fukugita$^{b}$, S. Hashimoto$^{c}$, N.
          Ishizuka$^{a,d}$, Y. Iwasaki$^{a,d}$,\\
          K. Kanaya$^{a,d}$,
 	 Y. Kuramashi$^{e}$, H. Mino$^{f}$, M. Okawa$^{e}$,
	A. Ukawa$^{a}$, T. Yoshi\'e$^{a,d}$\\
\\
        $^{a}$Institute of Physics,  University of Tsukuba\\
        Tsukuba, Ibaraki 305, Japan\\ \\
        $^{b}$Institute for Cosmic Ray Research, University of Tokyo\\
        Tanashi, Tokyo 188, Japan\\ \\
        $^{c}$Computing Research Center\\
        High Energy Accelerator Research Organization (KEK)\\
        Tsukuba, Ibaraki 305, Japan\\ \\
        $^{d}$Center for Computational Physics\\
        University of Tsukuba, Tsukuba, Ibaraki 305, Japan\\ \\
	$^{e}$Institute of Particle and Nuclear Studies\\
        High Energy Accelerator Research Organization (KEK)\\
        Tsukuba, Ibaraki 305, Japan\\ \\
        $^{f}$Faculty of Engineering, Yamanashi University\\
        Kofu 400, Japan\\
}

\date{}

\maketitle
\newpage
\begin{abstract}
We report on a study of two-flavor finite-temperature chiral phase
transition employing the Kogut-Susskind quark action and the
plaquette gluon action in lattice QCD for a lattice with $N_t=4$ temporal
size.  Hybrid R simulations of $10^4$ trajectories are made at quark masses of
$m_q=0.075, 0.0375, 0.02, 0.01$ in lattice units  for the  spatial sizes $8^3,
12^3$ and $16^3$.  The spatial size dependence of various
susceptibilities confirm the
previous conclusion of the absence of a phase transition down to $m_q=0.02$. At
$m_q=0.01$ an increase of susceptibilities is observed up to the largest
volume $16^3$ explored in the present work. We argue, however,
that this increase is likely to be due to an artifact of too small a
lattice size and it cannot be taken to be the evidence for
a first-order transition.  Analysis of critical exponents estimated from
the quark mass dependence of susceptibilities shows that they
satisfy hyperscaling consistent with a
second-order transition located at
$m_q=0$.  The exponents obtained from larger lattice, however,
deviate significantly
from both those of O(2),
which is the exact symmetry group of the Kogut-Susskind
action at finite lattice
spacing, and those of O(4) expected from an effective sigma model
analysis in the continuum limit.
\end{abstract}
\newpage

\section{Introduction}

The nature of the finite-temperature chiral phase transition has been
pursued using lattice QCD over many years.  The commonly adopted
simplification is to approximate the real world with
$N_f$ flavors of degenerate quarks.
Theoretical arguments\cite{sigmamodel} based on an effective sigma model that
preserves chiral symmetry of QCD
suggest the order of the transition changing from first to probably second
as $N_f$ decreases from $N_f\geq 3$ to $N_f=2$,
which is a reasonably close approximation to reality.

Lattice QCD simulations with the Kogut-Susskind quark action have
shown that the transition is indeed of first order for
$N_f=4$\cite{reviews,mtcfour,columbiafour,founf2}.
There are indications, though much less extensive,
that the $N_f=3$ transition is also of first order\cite{columbianf2}
in agreement with the theoretical expectation.
A physically important case of $N_f=2$, on the other hand,
has turned out to be more elusive.
In Ref.~\cite{founf2} a finite-size scaling analysis was attempted
at quark masses $m_q=0.025$ and 0.0125 in lattice units for a temporal
lattice size of
$N_t=4$ employing the spatial sizes $6^3$, $8^3$ and $12^3$.  While the
results clearly confirmed that the $N_f=2$ transition is much weaker than
that for $N_f=4$, a first-order transition could not be quite excluded
since various susceptibilities exhibited some
increase with spatial volume up to
$12^3$. Simulations on a  $16^3\times 4$ lattice at
$m_q=0.025$ and 0.01 carried out by the Columbia group\cite{columbianf2}
showed,
however, that the susceptibilities flatten off between $12^3$ and $16^3$
spatial sizes.  The combined results led to the conclusion
that a phase transition is absent down to the quark mass of
$m_q\approx 0.01-0.0125$, and this was taken to be
consistent with the prediction of
the  sigma model analysis\cite{sigmamodel} for a second-order transition
which takes place at $m_q=0$, and changes into a crossover at $m_q\ne 0$.

If the $N_f=2$ transition is indeed to follow the prediction of the effective
sigma model, critical exponents for the  $N_f=2$ system should agree with
those of the O(4) Heisenberg model in three dimensions. This point was first
studied by
Karsch\cite{karsch}.  Examining world data for the critical coupling
$\beta_c(m_q)$ as a function of quark mass $m_q$, he concluded
that the dependence is
consistent with a second-order scaling behavior with the O(4) critical
exponents.  This analysis has been extended in Ref.~\cite{karschlaermann} in
which various susceptibilities were measured on an $8^3\times 4$ lattice at
$m_q=0.075, 0.0375, 0.02$, and critical exponents were extracted from the quark
mass dependence of the peak height of susceptibilities. The results showed
that the magnetic exponent was in fair agreement with the O(4) value,
while that for the thermal exponent exhibited a sizable deviation.

In these earlier analyses
there are a number of respects that deserve further investigations.
First, the conclusion on the absence of a first-order transition at
$m_q\approx 0.01$
from finite size scaling was based on a combination of finite-size data from
two groups\cite{founf2,columbianf2}
which employed slightly different quark masses
($m_q=0.0125$\cite{founf2} and $0.01$\cite{columbianf2}).
There is also a suspicion that the simulation may not be long enough.
It is clearly desirable to reexamine
finite-size scaling behavior with a homogeneous data set generated
under the same simulation conditions.
Second, the method of second-order scaling analysis
should be applicable only for sufficiently large lattice sizes
to avoid finite-size effects. It is not clear if the spatial size of $8^3$
employed by
Karsch and Laermann\cite{karschlaermann} is sufficient,
especially toward light quark masses. Additional question is
whether the range of quark mass
$m_q=0.075-0.02$ they explored is small enough for the true critical behavior
to manifest in the susceptibilities. Thus an extension of their work toward
larger spatial sizes and smaller quark masses is undoubtfully desired.

In order to address these points we have carried out new
simulations for the two-flavor chiral phase transition with the
Kogut-Susskind quark action, and systematically collected data over a range of
spatial sizes and quark masses with statistics higher than in the
previous work.
Our simulations have been made for $m_q=0.075$, $0.0375$, $0.02$ and $0.01$
on lattices of size $8^3\times 4$, $12^3\times 4$ and
$16^3\times 4$, accumulating  10000 trajectories of the hybrid R algorithm for
each parameter set. In this article we present details of the runs and results
of our analyses on both finite-size and second-order scaling behavior.
The calculations have been performed on the Fujitsu VPP500/80 supercomputer
at KEK.

A preliminary account of our results was reported in Ref.~\cite{preliminary}.
A similar study has been carried out by the Bielefeld
group\cite{bielefeldchiral}
with lattice sizes up to $16^3\times 4$ but keeping
the quark mass only to $m_q\geq 0.02$.
The MILC Collaboration has recently started
simulations for small quark masses down to $m_q=0.008$ employing lattices as
large as $24^3$\cite{milc24}.

In Sec.~2 we describe details of our simulation.  In particular we explain our
method for computing the disconnected part of fermionic susceptibilities
which is non-trivial.  In Sec.~3 we discuss
finite-size scaling analysis for a given quark mass.
In Sec.~4 analyses of exponents and
scaling functions extracted from the quark mass dependence of susceptibilities
are presented. Conclusions of the present work are given in Sec.~5.

\section{Simulation and measurements}

\subsection{Simulation algorithm}

The present study is carried out with the plaquette action for gluons and the
Kogut-Susskind action for quarks.  The effective action is given by
\be
S_{eff}=-\frac{\beta}{6}\sum_{plaquette}\mbox{Tr}\left(U_{plaquette}\right)
	      -\frac{N_f}{4}\mbox{Tr}\log \left(D(U)^\dagger D(U)\right)_e,
\label{eq:action}
\ee
where $N_f=2$, the subscript $e$ means the even part,
and $D(U)$ denotes the Kogut-Susskind quark operator,
\be
D(U)=m_q+\frac{1}{2}\sum_\mu D_\mu (U)
\ee
with
\be
D_\mu (U)_{x,y}=\eta_{x\mu}
    \left(U_{x\mu}\delta_{x+\hat\mu,y}U_{x\mu}-
                  \delta_{x-\hat\mu,y}U_{y\mu}^\dagger\right).
\label{eq:deemu}
\ee

We employ the standard hybrid R algorithm to simulate the
system, adopting the same normalization of the step size $\Delta\tau$
as in the original literature\cite{hybridR}.
In the leap-frog update to solve the molecular dynamics
equations, link variables are assigned to half-integer time steps and conjugate
momenta to integer times.
Inversion of the quark operator is made with the
conjugate gradient algorithm.

\subsection{Observables and method of measurement}

We consider local observables defined by
\ben
\overline{\psi}\psi&=&\frac{1}{V}\sum_x\overline{\psi}_x\psi_x,\\
\overline{\psi}D_0\psi&=&\frac{1}{V}\sum_{x,y}
\overline{\psi}_xD_{0x,y}\psi_y,\\
P_\tau&=&\frac{1}{9V}\sum_{x}\sum_{1\leq i\leq3}
\mbox{Tr}\left(U_{x4i}\right),\\
P_\sigma&=&\frac{1}{9V}\sum_{x}\sum_{1\leq i<j\leq 3}
\mbox{Tr}\left(U_{xij}\right),\\
\Omega&=&\frac{1}{L^3}\sum_{\vec x}\Omega_{\vec x},\qquad
\Omega_{\vec x}=\frac{1}{3}\mbox{Tr}\left(\prod_{x_t=1}^{N_t}U_{x4}\right),
\een
where $V=L^3\cdot N_t$ denotes the lattice volume of an $L^3\times N_t$
lattice, $D_0$ the temporal hopping term of the Kogut-Susskind
operator as defined in (\ref{eq:deemu}),
and $U_{x\mu\nu}$ the plaquette in the $\mu\nu$ plane. In the course of
our simulation, we measure these quantities and the corresponding
susceptibilities given by
\ben
\chi_m&=&V\left[\langle\left(\overline{\psi}\psi\right)^2\rangle-
\langle\overline{\psi}\psi\rangle^2\right],\\
\chi_{t,f}&=&V\left[\langle \left(\overline{\psi}\psi\right)
                            \left(\overline{\psi}D_0\psi\right)\rangle
             -\langle\overline{\psi}\psi\rangle
                            \langle\overline{\psi}D_0\psi\rangle\right],\\
\chi_{t,i}&=&V\left[\langle \left(\overline{\psi}\psi\right)P_i\rangle
             -\langle\overline{\psi}\psi\rangle
                            \langle P_i\rangle\right],\qquad i=\sigma, \tau,\\
\chi_{e,f}&=&V\left[\langle\left(\overline{\psi}D_0\psi\right)^2\rangle-
\langle\overline{\psi}D_0\psi\rangle^2\right],\\
\chi_{e,i}&=&V\left[\langle \left(\overline{\psi}D_0\psi\right)P_i\rangle
             -\langle\overline{\psi}D_0\psi\rangle
                            \langle P_i\rangle\right],\qquad i=\sigma, \tau,\\
\chi_{e,ij}&=&V\left[\langle P_iP_j\rangle
            -\langle P_i\rangle\langle P_j\rangle\right]
        ,\qquad i,j=\sigma, \tau,\\
\chi_\Omega&=&V\left[\langle \Omega^2\rangle -\langle\Omega\rangle^2\right].
\een

Calculation of the fermionic susceptibilities $\chi_m$, $\chi_{t,f}$ and
$\chi_{e,f}$ is  non-trivial because of the presence of disconnected quark loop
contributions.  Let us illustrate our procedure for $\chi_m$.
After quark contractions and correcting by powers of $N_f/4$ for
normalization to $N_f$ flavors, $\chi_m$ is written
\ben
\chi_m&=&\chi_{disc}+\chi_{conn},\\
\chi_{disc}&=&\left(\frac{N_f}{4}\right)^2\frac{1}{V}\left[
\langle\left(\mbox{Tr}D^{-1}\right)^2\rangle-\langle\mbox{Tr}D^{-1}\rangle^2
\right],\\
\chi_{conn}&=&-\frac{N_f}{4}\frac{1}{V}\sum_{x,y}\langle
D_{x,y}^{-1}D_{y,x}^{-1}\rangle.
\een
We employ the volume source method without gauge fixing\cite{kuramashi} to
evaluate the two parts.  Let
\be
G_x^{a,b}\equiv\sum_y \left(D^{-1}\right)_{x,y}^{a,b}
\ee
be the quark propagator for unit source placed at every space-time site with a
given color $b$.  Define
\ben
O_1&=&\sum_{x,y}\sum_{a,b}G_x^{a,a}G_y^{b,b},\label{eq:volumeone}\\
O_2&=&\sum_{x,y}\sum_{a,b}G_x^{a,b}G_y^{b,a},\\
O_3&=&\sum_{x}\sum_{a,b}G_x^{a,a}G_x^{b,b},\\
O_4&=&\sum_{x}\sum_{a,b}G_x^{a,b}G_x^{b,a}.\label{eq:volumetwo}
\een
Up to terms which are gauge non-invariant, and hence vanish on the average, we
find
\ben
\left(\mbox{Tr}D^{-1}\right)^2&=&+\frac{9}{8}O_1-\frac{3}{8}O_2
                        -\frac{1}{8}O_3+\frac{3}{8}O_4,\\
\sum_{x,y}D_{x,y}^{-1}D_{y,x}^{-1}&=&-\frac{3}{8}O_1+\frac{9}{8}O_2
                        +\frac{3}{8}O_3-\frac{1}{8}O_4.
\een
Note that $O_1$ contains the connected contribution in addition to the
dominant disconnected part, and {\it vice versa} for $O_2$.  The
terms $O_3$ and $O_4$ represent contact contributions in
which the source and sink points of quark coincide.

To calculate the susceptibility $\chi_{t,f}$
we need to replace one of the volume
source propagator $G_x^{a,b}$ in (\ref{eq:volumeone}--\ref{eq:volumetwo})
by $(D_0G)_x^{a,b}$.  Both propagators should be replaced in this way
for $\chi_{e,f}$.

\subsection{Choice of run parameters}
\label{sec:parameter}

The distribution of observables generated by the hybrid R algorithm suffers
from systematic errors arising from a finite
molecular dynamics step size $\Delta\tau$ and a finite
stopping condition taken for
the conjugate gradient inversion of the quark operator.
For analyses of critical properties of phase transitions, potential
problems caused by these systematic errors are a shift of the critical
coupling $\beta_c$ and a modification of susceptibilities,
in particular a change in the
magnitude of the peak height of susceptibilities at $\beta=\beta_c$.  In order
to examine these effects we carry out test runs on an $8^3\times 4$
lattice at $m_q=0.02$.

\begin{figure*}
\centerline{\epsfxsize=90mm \epsfbox{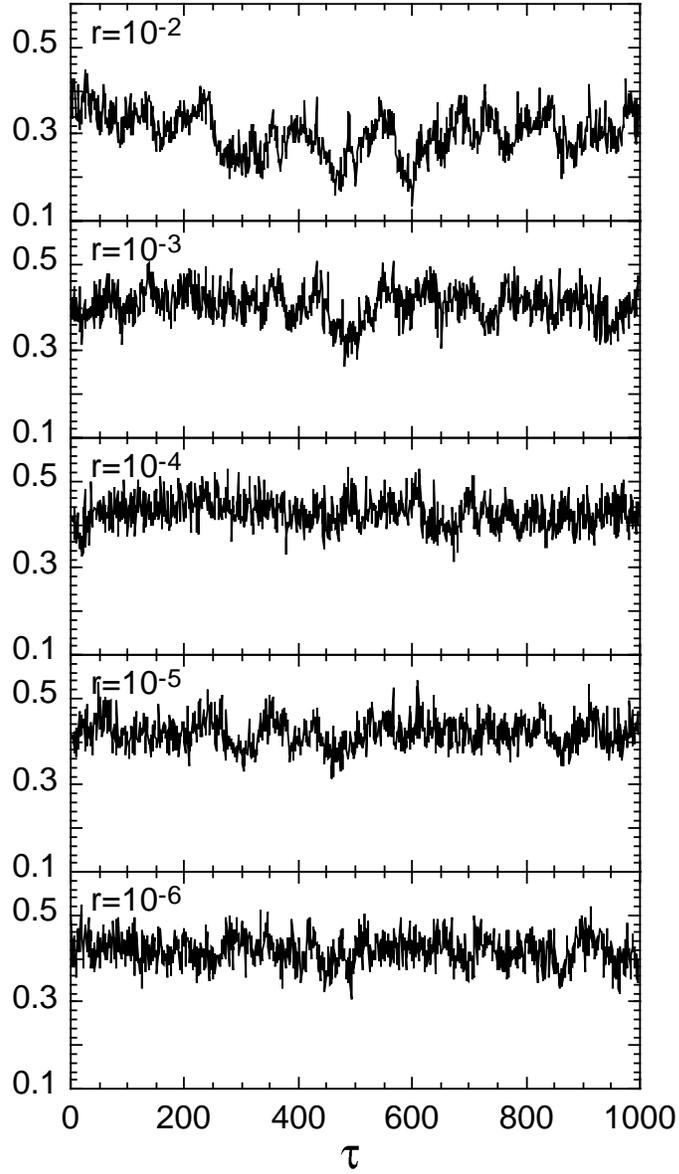}}
\caption{Time history of chiral order parameter
for a series of values of the stopping condition on an $8^3\times 4$
lattice at $m_q=0.02$ and $\Delta\tau=0.03$.}
\label{fig:cgtest}
\end{figure*}

Let us define a residual $r$ of the conjugate gradient
inversion algorithm applied for a source vector $b$ by
\be
r\equiv \sqrt{\frac{\vert\vert b_e-(D^\dagger Dx)_e\vert\vert^2}{3V}},
\ee
where, as in (\ref{eq:action}), the subscript $e$ means the even part.
The choice of the factor $3V$ is motivated by the fact that the norm of the
residue vector $\vert\vert b_e-(D^\dagger Dx)_e\vert\vert^2$ is proportional to
$V=L^3\cdot N_t$ for a Gaussian noise source employed for the hybrid R
algorithm.

To test effects of the stopping condition, we choose an approximate
value of the critical coupling
$\beta=5.282$ for $m_q=0.02$,
and generate 1000 trajectories of unit length with a fixed step
size of
$\Delta\tau=0.03$, varying the stopping condition from
$r=10^{-2}$ to $10^{-6}$.  The time histories of the chiral order parameter
$\overline{\psi}\psi$ for the runs are shown in
Fig~\ref{fig:cgtest}.  We observe that a looser stopping condition leads to
a smaller value and a larger magnitude of fluctuations of
$\overline{\psi}\psi$.
The results, however, are stable for $r\simlt 10^{-4}-10^{-5}$.
In all of our production runs we therefore take the condition given by
\be
r<10^{-6}.
\ee

Another possible measure of the stopping condition is to employ
the ratio
\be
\tilde r=\frac{\vert\vert b_e-(D^\dagger Dx)_e\vert\vert}
{\vert\vert x_e\vert\vert}.
\label{eq:stopalt}
\ee
With a Gaussian noise source we expect
$\vert\vert x_e\vert\vert^2\propto c(m_q)V$ with the coefficient $c(m_q)$
increasing as $m_q$ becomes smaller.  Thus our stopping condition is
relatively tighter
toward smaller quark masses compared to that given in terms of
(\ref{eq:stopalt}).

We examine systematic effects of the step size by carrying out runs of
$10000$ trajectories of unit length for the combinations $(\beta,
\Delta\tau)=(5.282, 0.01)$, $(5.284, 0.014)$, $(5.284, 0.02)$
with the stopping condition fixed at $r=10^{-6}$.

\begin{figure*}[t]
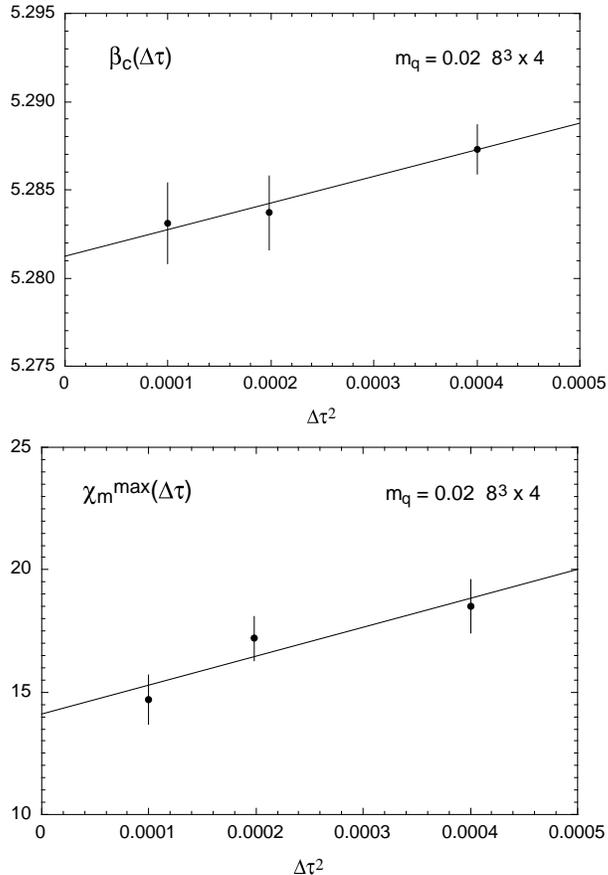

\centerline{\epsfxsize=80mm \epsfbox{betacvsdt.epsf}}
\centerline{\epsfxsize=80mm \epsfbox{chimmaxvsdt.epsf}}
\caption{(a) Critical coupling as a function of step size.
(b) Peak height of $\chi_m$ as a function of step size. Data are taken on
an $8^3\times 4$ lattice at $m_q=0.02$ with $r<10^{-6}$.}
\label{fig:stepsize}
\end{figure*}

The critical coupling
$\beta_c(\Delta\tau)$ estimated from the peak position of the chiral
susceptibility $\chi_m$, and its peak height $\chi_m^{max}(\Delta\tau)$ are
plotted in Fig.~\ref{fig:stepsize}(a) and (b) as a function of $\Delta\tau^2$,
where the standard reweighting technique\cite{reweighting} is employed to
estimate $\beta_c(\Delta\tau)$ and $\chi_m^{max}(\Delta\tau)$. We observe that
the results are consistent with an $O(\Delta\tau^2)$ dependence theoretically
expected\cite{hybridR,columbianf8,milcthermo}.  Since quark mass is expected to
affect the systematic error in the combination
$(\Delta\tau/m_q)^2$, we parametrize $\beta_c(\Delta\tau)$ and
$\chi_m^{max}(\Delta\tau)$ in the  form $a(1+c(\Delta\tau/m_q)^2)$ and find
$a=5.2812(26)$, $c=0.0011(6)$ for
$\beta_c(\Delta\tau)$ and $a=14.1(1.2)$, $c=0.34(16)$ for
$\chi_m^{max}(\Delta\tau)$.  These
values suggest that choosing
$\Delta\tau\approx\frac{m_q}{2}$ leads to an accuracy of
0.03\% (or $0.0015$ in magnitude) for $\beta_c$ and 9\% for $\chi_m^{max}$.
We think these accuracies to be sufficient compared to our statistical errors,
and adopt $\Delta\tau\approx\frac{m_q}{2}$ for our production runs.

\subsection{Summary of runs}

We carry out runs for the temporal lattice size $N_t=4$ at the quark masses
$m_q=0.075, 0.0375, 0.02$ and $0.01$.  For each quark mass we employ three
spatial lattice sizes given by $L=8, 12$ and $16$.  For each set $(m_q, L)$
we choose a single value of $\beta$ close to the critical coupling, which is
selected by preliminary short runs, and carry out a long simulation of 10000
trajectories of unit length starting from an ordered configuration
using the stopping condition as described in Sec.~\ref{sec:parameter}.
Variation of observables as a function of $\beta$ is calculated by the
reweighting technique\cite{reweighting} from a single run.

In applying the reweighting technique one may consider
an alternative procedure of making a number of
shorter runs for a set of values of $\beta$ around $\beta_c$.
In practice we find long-range fluctuations of
$O(1000)$ trajectories  toward smaller quark masses and larger volumes,
so that a
simulation at a single parameter point is already quite computer time
intensive to get rid of these fluctuations.
We therefore adopt the approach of making a single long run at a well-chosen
value of $\beta$ in the present work.

In Table~\ref{tab:runs} we list the values of $\beta$ where our runs are
carried out and the molecular dynamics step size $\Delta\tau$ used.
Two runs are made
for $m_q=0.01$ on a $12^3\times 4$ lattice since the first run at $\beta=5.266$
turns out to be predominantly in the low-temperature phase.
We collect time histories of the chiral order parameter $\overline{\psi}\psi$
and their histograms in Fig~\ref{fig:history}.

\begin{figure*}
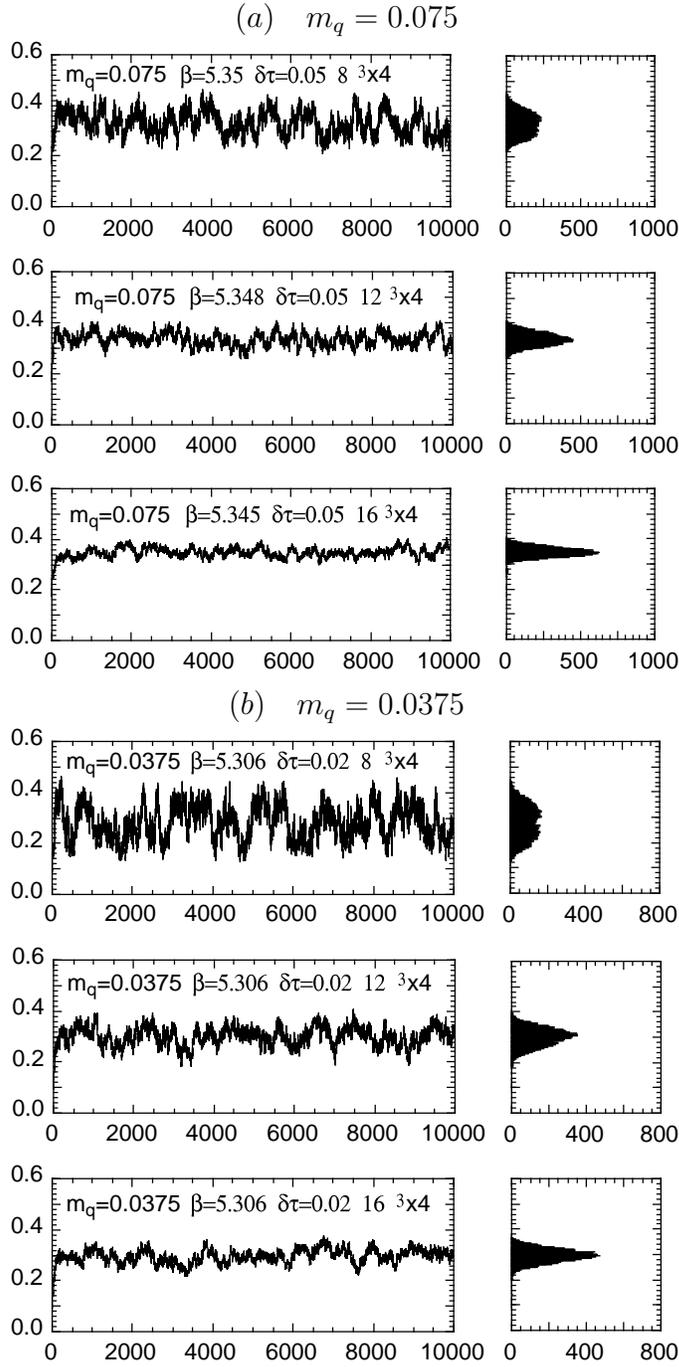

\centerline{$(a)\quad m_q=0.075$}
\centerline{\epsfxsize=90mm \epsfbox{history_a.epsf}}
\centerline{$(b)\quad m_q=0.0375$}
\centerline{\epsfxsize=90mm \epsfbox{history_b.epsf}}
\caption{Time history of $\overline{\psi}\psi$ and histograms of
the runs.}
\label{fig:history}
\end{figure*}

\setcounter{figure}{2}
\begin{figure*}
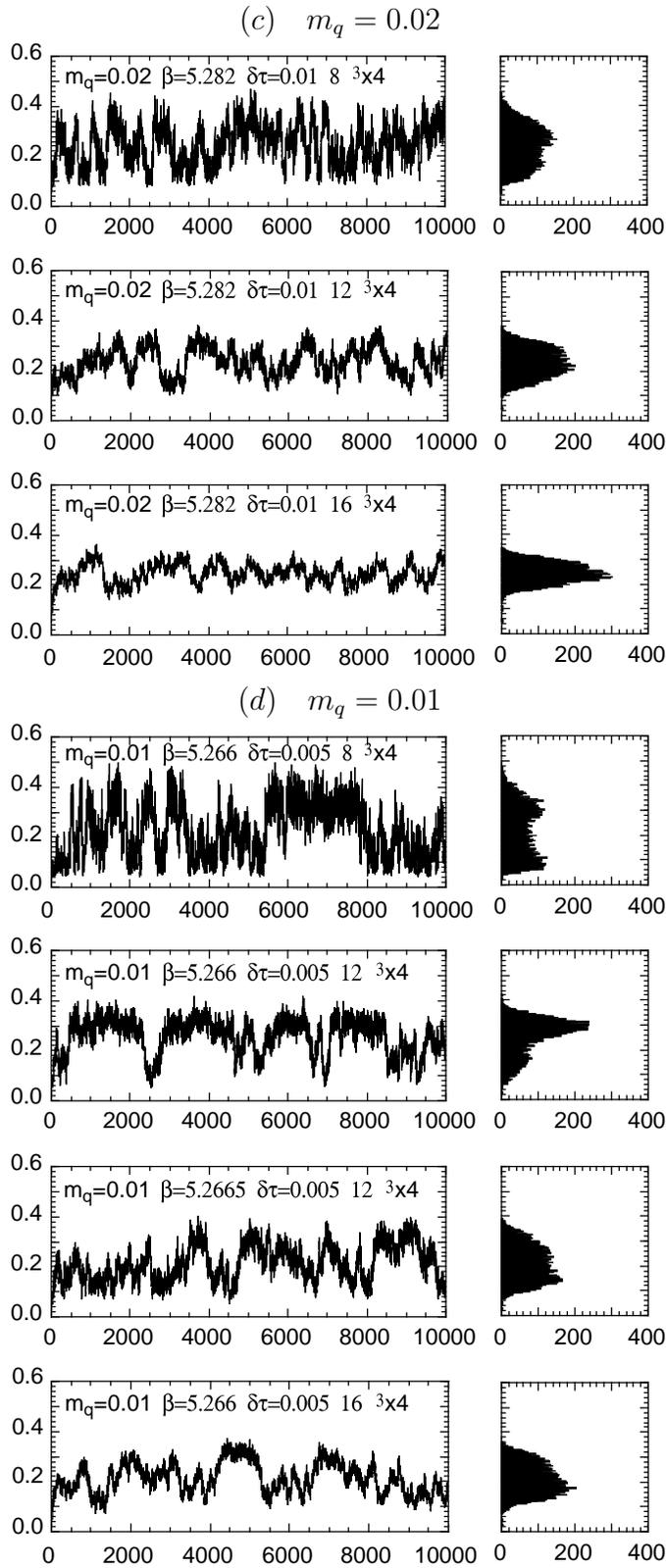

\centerline{$(c)\quad m_q=0.02$}
\centerline{\epsfxsize=90mm \epsfbox{history_c.epsf}}
\centerline{$(d)\quad m_q=0.01$}
\centerline{\epsfxsize=90mm \epsfbox{history_d.epsf}}
\caption{Continued}
\end{figure*}

In all of the runs observables are calculated at every trajectory,
discarding  the initial 2000
trajectories of each run.  Jackknife analysis is carried out for the
reminaing 8000 trajectories to estimate errors.  Examining the bin size
of 400 and 800 we find that the magnitude of errors is stable,
and we adopt 800 for the bin size of our error estimations.

\begin{table}
\begin{center}
\setlength{\tabcolsep}{1pc}
\caption{Parameters of our runs.  For each parameter point 10000 trajectories
are generated with the stopping condition $r=10^{-6}$ for the conjugate
gradient solver.}
\label{tab:runs}
\vspace*{10mm}
\begin{tabular}{lllll}
\hline
$N_s$&$m_q=0.075$&$0.0375$&$0.02$&$0.01$\\
     &$\Delta\tau=0.05$&0.02&0.01&0.005\\
\hline
8     &$\beta=5.35$&5.306&5.282&5.266\\
12    &$\beta=5.348$&5.306&5.282&5.266\\
      &     &     &     &5.2665\\
16    &$\beta=5.345$&5.306&5.282&5.266\\
\hline
\end{tabular}
\end{center}
\end{table}

\begin{table}
\begin{center}
\setlength{\tabcolsep}{0.2pc}
\caption{Peak position $\beta_c$ and peak height $\chi^{max}$ of
various susceptibilities for each quark mass $m_q$ and spatial lattice size
$L$.}
\label{tab:heights}
\vspace*{3mm}
\begin{tabular}{ccccccc}
\hline
       &\multicolumn{2}{c}{$L=8$} &\multicolumn{2}{c}{$L=12$}
&\multicolumn{2}{c}{$L=16$}\\
\hline
$m_q$&$\beta_c$&$\chi_m^{max}$&
$\beta_c$&$\chi_m^{max}$&$\beta_c$&$\chi_m^{max}$\\
\hline
0.0750	&5.3494(17)&5.9(0.3) &5.3477(14)&6.3(0.6)&5.3443(18)&6.3(0.7)	\\
0.0375	&5.3073(18)&10.5(0.5)&5.3099(16)&11.6(1.6)&5.3072(7)&14.1(2.2)\\
0.0200	&5.2831(23)&14.7(1.0)&5.2823(8)	&24.6(3.1)&5.2819(5)&22.9(2.2)	\\
0.0100	&5.2665(20)&24.4(1.6)&5.2681(7)	&44.4(6.2)&5.2657(4)&63.9(12.3)	\\
0.0100	&	&	&5.2665(6) &38.0(3.9)&	&			\\
\hline
$m_q$  &$\beta_c$      &$\chi_{t,f}^{max}$& $\beta_c$
&$\chi_{t,f}^{max}$&$\beta_c$      &$\chi_{t,f}^{max}$\\
\hline
0.0750	&5.3484(17)&-1.97(0.13)&5.3472(12)&-2.08(0.24)&5.3441(14)&-2.17(0.24)\\
0.0375	&5.3064(19)&-2.76(0.16)&5.3086(14)&-3.20(0.43)&5.3069(7)&-4.11(0.72)\\
0.0200	&5.2821(18)&-3.23(0.27)&5.2819(9)&-5.82(0.77)&5.2817(5)	&-5.71(0.56)\\
0.0100	&5.2658(20)&-4.77(0.39)&5.2678(7)&-9.02(1.31)&5.2655(4)	&-14.05(3.25)\\
0.0100	&	&	&5.2661(7)&-8.20(0.89)&		&		\\
\hline
$m_q$  &$\beta_c$      &$\chi_{t,\sigma}^{max}$& $\beta_c$
&$\chi_{t,\sigma}^{max}$&$\beta_c$      &$\chi_{t,\sigma}^{max}$\\
\hline
0.0750	&5.3483(16)&-0.71(0.05)&5.3474(14)&-0.76(0.10)&5.3441(16)&-0.80(0.11)\\
0.0375	&5.3060(19)&-1.04(0.07)&5.3085(19)&-1.15(0.17)&5.3069(8)&-1.48(0.27)\\
0.0200	&5.2816(22)&-1.24(0.11)&5.2819(9)&-2.22(0.29)&5.2816(6)&-2.12(0.22)\\
0.0100	&5.2656(22)&-1.91(0.15)&5.2679(7)&-3.70(0.57)&5.2656(4)&-5.62(1.17)\\
0.0100	&	&	&5.2661(7)&-3.18(0.39)&		&	\\
\hline
$m_q$  &$\beta_c$      &$\chi_{t,\tau}^{max}$& $\beta_c$
&$\chi_{t,\tau}^{max}$&$\beta_c$      &$\chi_{t,\tau}^{max}$\\
\hline
0.0750	&5.3482(16)&-0.79(0.06)&5.3473(14)&-0.84(0.11)&5.3439(17)&-0.89(0.12)\\
0.0375	&5.3061(19)&-1.16(0.08)&5.3088(22)&-1.29(0.22)&5.3069(8)&-1.68(0.32)\\
0.0200	&5.2817(22)&-1.40(0.12)&5.2819(9)&-2.53(0.33)&5.2816(6)&-2.41(0.25)\\
0.0100	&5.2656(21)&-2.12(0.17)&5.2679(7)&-4.09(0.63)&5.2656(4)&-6.32(1.35)\\
0.0100	&	&	&5.2661(7)&-3.56(0.42)&		&		\\
\hline
$m_q$  &$\beta_c$      &$\chi_{e,f}^{max}$& $\beta_c$
&$\chi_{e,f}^{max}$&$\beta_c$      &$\chi_{e,f}^{max}$\\
\hline
0.0750	&5.3480(17)&1.31(0.06)&5.3462(16)&1.41(0.10)&5.3441(13)&1.38(0.09)\\
0.0375	&5.3062(25)&1.53(0.06)&5.3078(15)&1.55(0.11)&5.3065(7)&1.96(0.25)\\
0.0200	&5.2825(19)&1.60(0.11)&5.2813(13)&2.09(0.18)&5.2815(10)&2.18(0.18)\\
0.0100	&5.2626(31)&2.14(0.19)&5.2676(9)&2.39(0.34)&5.2656(4)&3.82(0.84)\\
0.0100	&	&	&5.2657(6)&2.58(0.31)&		&		\\
\hline
$m_q$  &$\beta_c$      &$\chi_{e,\sigma}^{max}$& $\beta_c$
&$\chi_{e,\sigma}^{max}$&$\beta_c$      &$\chi_{e,\sigma}^{max}$\\
\hline
0.0750&5.3478(17)&0.236(0.021)&5.3470(14)&0.259(0.040)&5.3441(15)&
0.271(0.041)\\
0.0375&5.3058(21)&0.276(0.021)&5.3080(18)&0.321(0.050)&5.3067(8)&0.438(0.088)\\
0.0200&5.2813(20)&0.279(0.031)&5.2816(9)&0.538(0.076)&5.2815(7)&0.530(0.059)\\
0.0100&5.2652(21)&0.399(0.032)&5.2677(7)&0.754(0.120)&5.2655(4)&1.245(0.314)\\
0.0100&		&	&5.2658(7)&0.715(0.089)&	&	\\
\hline
\end{tabular}
\end{center}
\end{table}
\setcounter{table}{1}
\begin{table}
\begin{center}
\setlength{\tabcolsep}{0.2pc}
\caption{Continued}
\vspace*{3mm}
\begin{tabular}{ccccccc}
\hline
$m_q$  &$\beta_c$      &$\chi_{e,\tau}^{max}$& $\beta_c$
&$\chi_{e,\tau}^{max}$&$\beta_c$      &$\chi_{e,\tau}^{max}$\\
\hline
0.0750	&5.3477(17)&0.302(0.025)&5.3470(14)&0.321(0.043)&5.3439(16)&
0.335(0.045)
\\
0.0375&5.3059(20)&0.351(0.023)&5.3082(19)&0.399(0.060)&5.3067(8)&0.539(0.100)\\
0.0200&5.2812(19)&0.358(0.032)&5.2816(9)&0.659(0.085)&5.2815(7)&0.642(0.066)\\
0.0100&5.2652(21)&0.487(0.039)&5.2677(7)&0.871(0.133)&5.2655(4)&1.440(0.362)\\
0.0100&		&	&5.2658(7)&0.834(0.098)&	&	\\
\hline
$m_q$  &$\beta_c$      &$\chi_{e,\sigma\sigma}^{max}$& $\beta_c$
&$\chi_{e,\sigma\sigma}^{max}$&$\beta_c$      &$\chi_{e,\sigma\sigma}^{max}$\\
\hline
0.0750&5.3478(15)&0.141(0.008)&5.3474(14)&0.149(0.015)&5.3443(14)&
0.159(0.019)\\
0.0375&
5.3053(21)&0.161(0.010)&5.3075(22)&0.170(0.020)&5.3067(9)&0.212(0.033)\\
0.0200&5.2808(23)&0.162(0.013)&5.2816(9)&0.261(0.028)&5.2814(7)&0.252(0.023)\\
0.0100&	5.2650(22)&0.214(0.016)&5.2678(8)&0.369(0.053)&5.2655(5)&0.557(0.111)\\
0.0100&	&	&5.2658(7)&0.331(0.040)&	&	\\
\hline
$m_q$  &$\beta_c$      &$\chi_{e,\sigma\tau}^{max}$& $\beta_c$
&$\chi_{e,\sigma\tau}^{max}$&$\beta_c$      &$\chi_{e,\sigma\tau}^{max}$\\
\hline
0.0750&5.3479(15)&0.135(0.010)&5.3474(14)&0.143(0.018)&5.3441(16)&
0.152(0.021)\\
0.0375&5.3055(21)&0.155(0.010)&5.3079(24)&0.166(0.024)&5.3067(9)&0.218(0.039)\\
0.0200&5.2809(23)&0.157(0.014)&5.2816(10)&0.272(0.032)&5.2814(6)&0.262(0.026)\\
0.0100&5.2651(22)&0.214(0.018)&5.2678(7)&0.385(0.058)&5.2655(5)&0.600(0.127)\\
0.0100&		&	&5.2658(7)&0.345(0.043)&	&	\\
\hline
$m_q$  &$\beta_c$      &$\chi_{e,\tau\tau}^{max}$& $\beta_c$
&$\chi_{e,\tau\tau}^{max}$&$\beta_c$      &$\chi_{e,\tau\tau}^{max}$\\
\hline
0.0750&5.3477(16)&0.168(0.012)&5.3473(13)&0.175(0.020)&5.3438(18)&
0.186(0.024)\\
0.0375&5.3055(20)&0.189(0.011)&5.3083(27)&0.202(0.030)&5.3067(9)&0.265(0.046)\\
0.0200&5.2809(22)&0.194(0.015)&5.2816(10)&0.326(0.036)&5.2814(6)&0.314(0.030)\\
0.0100&5.2650(22)&0.254(0.019)&5.2678(7)&0.442(0.065)&5.2655(5)&0.689(0.146)\\
0.0100&		&	&5.2659(7)&0.403(0.047)&	&	\\
\hline
$m_q$  &$\beta_c$      &$\chi_{\Omega}^{max}$& $\beta_c$
&$\chi_{\Omega}^{max}$&$\beta_c$      &$\chi_{\Omega}^{max}$\\
\hline
0.0750&5.3479(19)&4.21(0.28)&5.3462(13)&4.52(0.51)&5.3439(14)&4.63(0.54)\\
0.0375&5.3064(20)&4.10(0.25)&5.3087(15)&4.53(0.57)&5.3068(7)&6.05(1.03)	\\
0.0200&5.2823(20)&3.73(0.32)&5.2819(9)&6.30(0.75)&5.2817(5)&6.17(0.62)	\\
0.0100&5.2655(22)&4.45(0.34)&5.2677(8)&8.06(1.17)&5.2655(4)&12.57(3.00)	\\
0.0100&	&	&5.2660(7)&7.21(0.78)&		&		\\
\hline
\end{tabular}
\end{center}
\end{table}

\section{Analysis of spatial volume dependence}

\subsection{Finite-size scaling analysis}

\begin{figure*}
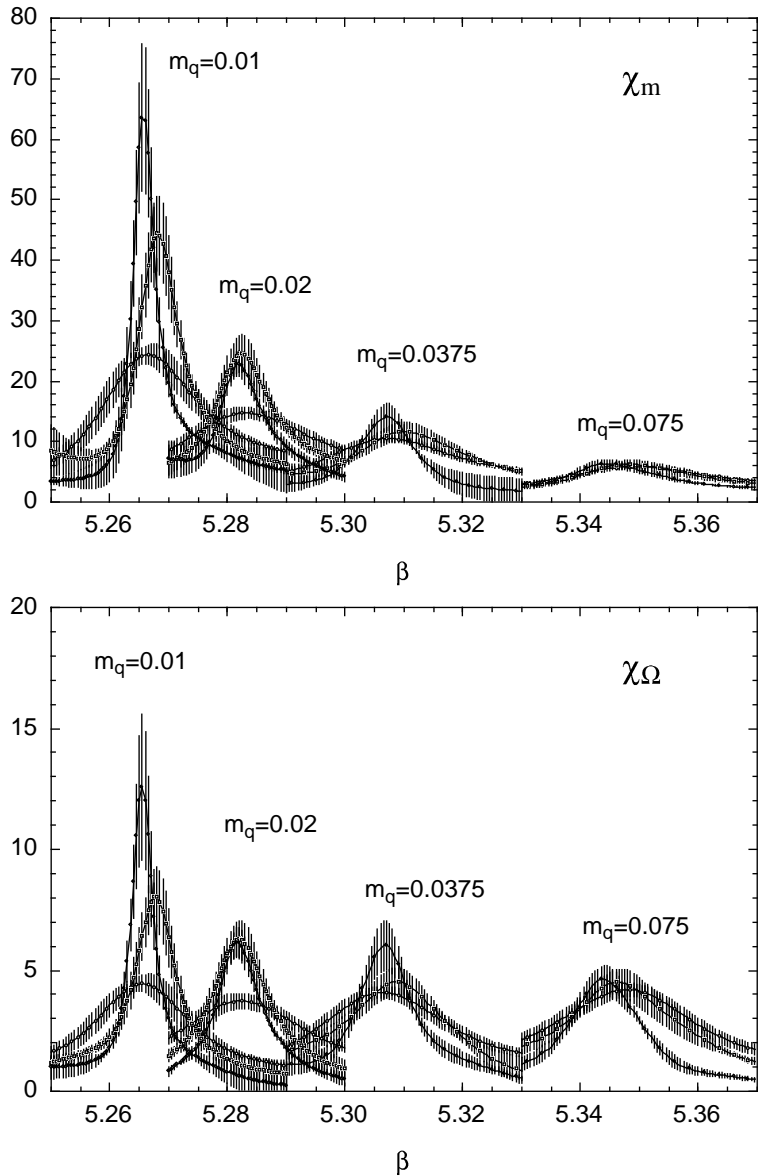

\centerline{\epsfxsize=100mm \epsfbox{shape-a.epsf}}
\centerline{\epsfxsize=100mm \epsfbox{shape-b.epsf}}
\caption{(a) Chiral susceptibility $\chi_m$ as a function of $\beta$.
(b) Same for the Polyakov susceptibility $\chi_\Omega$.  For $L=12$ at
$m_q=0.01$ the run with $\beta=5.266$ is employed. }
\label{fig:volume}
\end{figure*}

\begin{figure*}
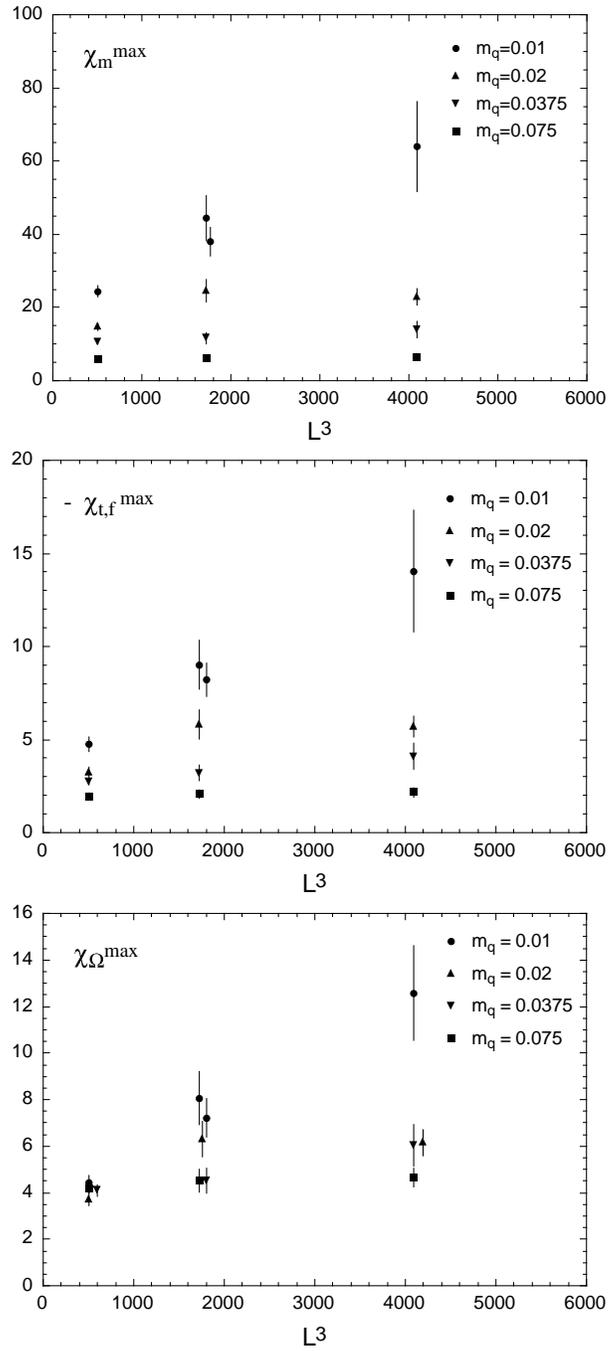

\centerline{\epsfxsize=80mm \epsfbox{heightvsvolume-a.epsf}}
\centerline{\epsfxsize=80mm \epsfbox{heightvsvolume-b.epsf}}
\centerline{\epsfxsize=80mm \epsfbox{heightvsvolume-c.epsf}}
\caption{Peak height of the susceptibility $\chi_m$, $\chi_{t,f}$ and
$\chi_\Omega$ as a function of spatial volume $L^3$.}
\label{fig:heightvsvolume}
\end{figure*}

We start our analysis with an examination of the spatial volume
dependence of susceptibilities for each quark mass.
The $\beta$ dependence of susceptibilities, evaluated with the
reweighting technique for each $m_q$ and spatial size $L$, is illustrated for
the chiral and Polyakov susceptibilities,
$\chi_m$ and $\chi_\Omega$, in Fig~\ref{fig:volume}.  Let us denote by
$\beta_c$ and $\chi^{max}$ the position and height of the peak of a
susceptibility.
Our numerical results for these quantities are summarized in
Table~\ref{tab:heights} for each of the susceptibilities defined by
(9--15).  As typical examples, we plot $\chi_m^{max}$, $\chi_{t,f}^{max}$ and
$\chi_\Omega^{max}$  as a function of spatial volume $L^3$ in
Fig~\ref{fig:heightvsvolume}.  Two points for $m_q=0.01$ on a $12^3$
lattice represent two runs at these parameters.  The agreement between
the two points justifies the robustness of the reweighting method:
the method works well even if the simulation is carried out in
one side of the two phases, i.e., at $\beta$ off the critical value.
The behavior of other susceptibilities are
similar as one may find from Table~\ref{tab:heights}.

For the heavier quark masses of $m_q=0.075$ and 0.0375 the peak height of the
susceptibilities increases little over the sizes $L=8-16$, showing that
a phase transition is absent for these masses.
For $m_q=0.02$, a significant increase is seen between $L=8$ and 12.  The
increase, however, does not continue beyond $L=12$, with the peak height for
$L=16$ consistent with that for $L=12$.  We then conclude an absence of a phase
transition also for $m_q=0.02$.  The histogram shown in
Fig.~\ref{fig:history}(c) provides further support for this conclusion;
while the histogram for the size
$L=8$ is broad and even hint at a possible presence of a double peak structure,
such an indication for metastability is less visible for $L=12$, and a single
peak structure becomes quite manifest for $L=16$.

For $m_q=0.01$ the peak height also increases between $L=8$ and 12.
Furthermore, the
increase continues up to $L=16$. In fact the rate of increase is
consistent with a linear behavior in volume, which is expected for
a first-order phase transition.

We think, however, that caution must be exercised to draw a conclusion
solely from Fig~\ref{fig:heightvsvolume}.  Comparing the time histories of
$\overline{\psi}\psi$  for the three lattice sizes $L=8$, $12$ and $16$ in
Fig~\ref{fig:history}(d), we observe that a flip-flop behavior between two
different values of
$\overline{\psi}\psi$ is most distinct for the smallest lattice size $L=8$,
and the time histories for the larger lattice sizes $L=12$ and 16 are
dominated more by irregular patterns, the width of fluctuations
becoming smaller as the size increases.
These features are also reflected in the histograms.
A double-peak distribution, clearly seen for
$L=8$, is less evident for $L=12$ and barely visible for $L=16$.
Moreover, the width of the distribution is narrower for larger lattice
sizes. These trends show a marked contrast with the case of
the first-order deconfining phase transitions of the
pure SU(3) gauge theory and of four-flavor QCD, where
a flip-flop behavior in the time history
and a double-peak distribution in histograms become
progressively pronounced
toward larger spatial volumes, for instance,
as is seen in Fig.~1 of Ref.~\cite{foupuregauge}.

The observations above indicate that the increase of
susceptibilities seen for $m_q=0.01$ is due to insufficient spatial volume,
which is similar to an increase between $L=8$ and $12$ for $m_q=0.02$ for
which the susceptibilities level off for $L=16$.
In order to make a comparison of volume dependence for different
quark masses, we need to normalize the lattice
size in terms of a relevant length scale, which may be taken to be the pion
correlation length $\xi_\pi=1/m_\pi$ at zero temperature.
Results of $m_\pi$ precisely at the values of $\beta$ and
$m_q$ where our simulations are made are not available.
The MILC Collaboration, however, has given a parametrization of
available data for $\pi$ and $\rho$ meson masses as a function of
$\beta$ and $m_q$\cite{milcthermo}, from which we find
$\xi_\pi\approx 3.0$ for $m_q=0.02$ and $\xi_\pi\approx 4.4$ for $m_q=0.01$ at
the respective critical couplings.  Hence the size $L=8$ for $m_q=0.02$
roughly corresponds to $L=12$ for $m_q=0.01$, and $L=12$ to $L=16$.
Comparing the histograms for $m_q=0.02$ and  $0.01$ which are
in correspondence in this sense, we find that they are similar not only in
shape but also in the trend that a double peak type distribution changes
toward that of a single peak for larger sizes.

\begin{figure*}
\centerline{\epsfxsize=80mm \epsfbox{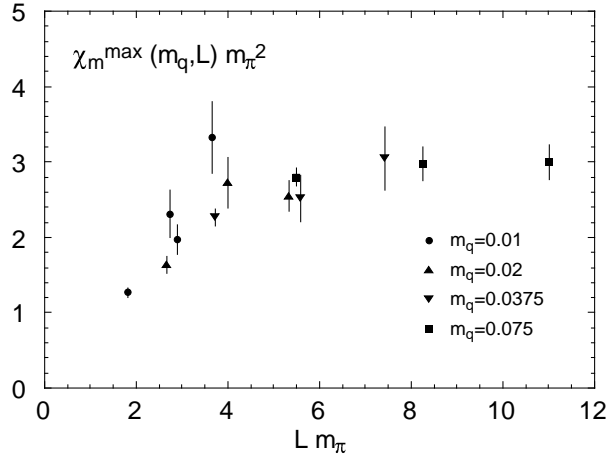}}
\caption{Peak height of chiral susceptibility $\chi_m$ as a function of
spatial size $L$, both normalized by zero-temperature pion mass $m_\pi$.}
\label{fig:heightvspivolume}
\end{figure*}

A more quantitative comparison is made in Fig.~\ref{fig:heightvspivolume}
where we plot the dimensionless combination $\chi_m^{max}\cdot m_\pi^2$
against $Lm_\pi$.  Data points for various
quark masses and spatial sizes roughly fall on a single curve,
and the increase observed up to $L=16$ for $m_q=0.01$
does not stand out as particularly large.
It is quite plausible that the peak height for
$m_q=0.01$ levels off if measured on a larger lattice, {\it e.g.,} $L=24$.

While a definitive conclusion has to await simulations on larger spatial
sizes, our examinations lead us to conclude that a first-order phase
transition is absent also at $m_q=0.01$.


\subsection{Comparison with previous studies}

Finite-size analyses similar to those reported here were previously carried
out by two groups\cite{founf2,columbianf2}.  In Ref.~\cite{founf2} runs of
$4000-10000$ trajectories of unit length were made for the spatial sizes
$6^3$, $8^3$ and $12^3$ at
$m_q=0.0125$ and $0.025$ using the step size of $\Delta\tau=0.02$ for both
cases.  In Ref.~\cite{columbianf2} a larger spatial lattice of $16^3$ was
employed, and 2500 trajectories were generated at $m_q=0.01
(\Delta\tau=0.0078)$ and $m_q=0.025 (\Delta\tau=0.01)$.
The quantities examined in these studies were the
Polyakov susceptibility $\chi_\Omega$ and the pseudo chiral susceptibility
defined by
\be
\chi_c\equiv V\left[\langle\left(\frac{1}{3}\xi^\dagger
D^{-1}\xi\right)^2\rangle
 -\langle\frac{1}{3}\xi^\dagger D^{-1}\xi\rangle^2\right],
\ee
where $\xi$ is a Gaussian noise vector.

\begin{figure*}
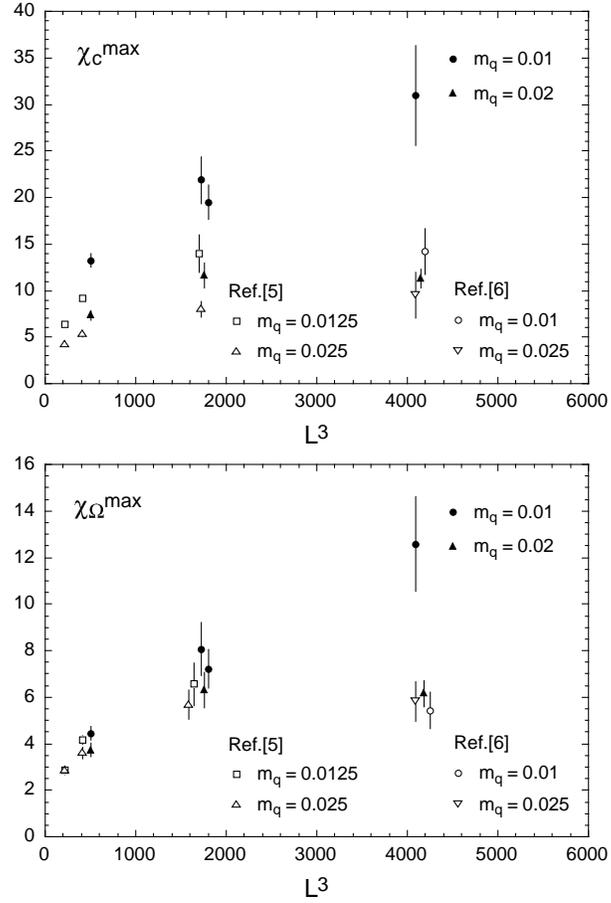

\centerline{\epsfxsize=80mm \epsfbox{olddata-a.epsf}}
\centerline{\epsfxsize=80mm \epsfbox{olddata-b.epsf}}
\caption{(a) Peak height of pseudo chiral susceptibility $\chi_c$ as a
function of $m_q$ for various spatial volumes.
(b) Same for the Polyakov susceptibility $\chi_\Omega$. Previous results
are plotted with open symbols.}
\label{fig:olddata}
\end{figure*}

We also calculate the pseudo chiral susceptibility in the present work.  In
Fig.~\ref{fig:olddata} (a)  previous data for this quantity  from
Refs.~\cite{founf2,columbianf2} are compared with the new results.  A similar
comparison for the Polyakov loop susceptibility is made in
Fig.~\ref{fig:olddata} (b). We observe that the data are consistent for the
sizes $L=8$ and $L=12$.  A reasonable agreement is also seen
between the present simulation and the earlier results for $L=16$ at
$m_q=0.025-0.02$.  At the smallest quark
mass of $m_q=0.01$, however, the result from Ref.~\cite{columbianf2}
is by a factor
$2-3$ smaller compared to our values.

A technical point to note in the calculation of Ref.~\cite{columbianf2}
for $\chi_c$ is that it used a multiple set of noise vectors for each
configuration in contrast to a single vector employed in Ref.~\cite{founf2}
and the present work.
This, however, would not be the main source of the discrepancy since the
result of Ref.~\cite{columbianf2} for $m_q=0.025$ is in
agreement with the other calculations.  This
difference also cannot explain the discrepancy in the Polyakov susceptibility.
We think it likely that the underestimate of
Ref.~\cite{columbianf2} originates
from a shorter length of their run.
Indeed dividing our full set of trajectories at $m_q=0.01$ and $L=16$
into subsets of 2500 each,
we find susceptibilities reduced by a similar factor for some of the subsets
owing to a long-range fluctuations over $\tau\sim O(1000)$.

\section{Analysis of quark mass dependence}

\subsection{Scaling laws and exponents}
\label{sec:scaling}

We have seen in the previous section that our finite-size data do not show
clear evidence for a first order phase transition down to $m_q=0.01$.  In the
present section we assume that the two-flavor chiral transition is of
second-order which takes place at $m_q=0$, and turns into a smooth crossover
for $m_q\ne 0$.
Various scaling laws follow from this assumption for the quark
mass dependence of the susceptibilities. We analyze to what extent our data
support the expected scaling laws.  In particular, we examine whether the
scaling exponents agree with the O(4) values as predicted by the
effective sigma model analysis\cite{sigmamodel}, or at least the O(2) values
corresponding to exact $U(1)$ chiral symmetry of the Kogut-Susskind
quark action used in our simulations.

The scaling laws follow from a well-known renormalization-group argument which
predicts that the leading singularity of the free energy per unit volume has
the
scaling form
\be
f_s(t,h)=h^{d/y_h}\phi(t\cdot h^{-y_t/y_h}),
\ee
where $t$ and $h$ are reduced temperature and quark mass, $y_t$ and $y_h$
are the thermal and magnetic critical exponents, and $d=3$ is
the space dimension.
We take the reduced variables to be
\ben
t&=&\beta_c(m_q)-\beta_c(0),\\
h&=&m_q,
\een
where $\beta_c(m_q)=6/g_c^2(m_q)$ denotes the pseudo critical coupling
defined as the peak
position of a susceptibility for a given quark mass $m_q$.  The choice for $h$
corresponds to $h\propto m_q/T$ up to a numerical factor of 4.  The scaling law
for the pseudo critical coupling is then given by
\be
\beta_c(m_q)=\beta_c(0)+c_gm_q^{z_g}
\label{eq:zg}
\ee
with
\be
z_g=\frac{y_t}{y_h}.
\label{eq:zge}
\ee
One can define three types of susceptibilities depending on the combination of
variables taken for the second derivative of the free energy.  The
$(h,h)$ combination corresponds to the chiral susceptibility $\chi_m$, and we
find the scaling form of its peak height to be
\be
\chi_m^{max}(m_q)=c_mm_q^{-z_m},
\label{eq:zm}
\ee
where
\be
z_m=2-\frac{d}{y_h}.
\ee
The $t$ derivative generates susceptibilities involving the energy operator
$\epsilon$.  Decomposing $\epsilon$ into the gluon terms
that depend on the spatial and temporal plaquette averages
and the quark term proportional to $\overline{\psi}D_0\psi$, we expect
\ben
\chi_{t,i}^{max}(m_q)&=&c_{t,i}\ m_q^{-z_{t,i}},\qquad i=f,\sigma,\tau ,\\
\chi_{e,i}^{max}(m_q)&=&c_{e,i}\ m_q^{-z_{e,i}},\qquad i=f,\sigma,\tau ,\\
\chi_{e,ij}^{max}(m_q)&=&c_{e,ij}\ m_q^{-z_{e,ij}},\qquad i,j=\sigma,\tau
\label{eq:zte}
\een
For these susceptibilities only the leading exponent needs to be constrained by
the thermal and magnetic exponents, {\it i.e.,}
\ben
z_t&=&1+\frac{y_t}{y_h}-\frac{d}{y_h},
\qquad z_t=\mbox{max}_{i=f,\sigma,\tau}\{z_{t,i}\},\\
z_e&=&\frac{2y_t}{y_h}-\frac{d}{y_h},
\qquad
z_e=\mbox{max}_{i=f,\sigma,\tau,j,k=\sigma,\tau}\{z_{e,i},z_{e,jk}\}.
\een
Since all the exponents are expressed in terms of $y_t$ and $y_h$,
two relations exist among the four exponents $z_g, z_m, z_t$ and $z_e$,
which may be taken to be
\ben
z_g+z_m&=&z_t+1,
\label{eq:consistencyone}\\
2z_t-z_m&=&z_e.
\label{eq:consistencytwo}
\een

\subsection{Results for exponents}
\label{sec:exponents}

Our study of exponents is based on the results for the  peak position and
height of various susceptibilities summarized in Table~\ref{tab:heights}.
For $m_q=0.01$ and $L=12$ two runs are made.  We present results employing the
first run carried out at $\beta=5.266$, since the exponents obtained
with the second run are consistent with those with the first run.

\begin{figure*}
\centerline{\epsfxsize=80mm \epsfbox{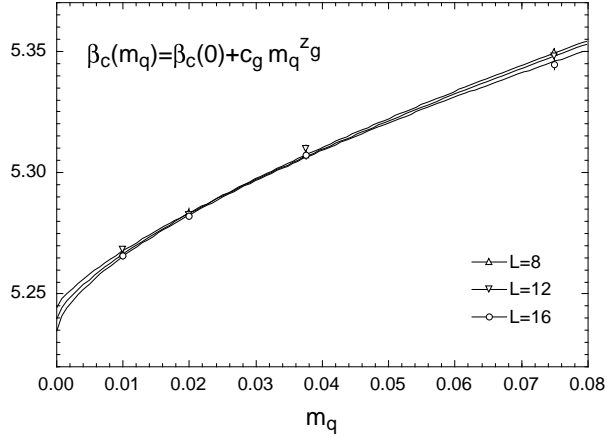}}
\caption{Pseudo critical coupling $\beta_c(m_q)$ as a function of $m_q$
determined from the peak position of susceptibility $\chi_m$.
Lines are fits to (\protect\ref{eq:zg}).}
\label{fig:criticalcoupling}
\end{figure*}

Let us start with an examination of the exponent $z_g$ that governs the scaling
behavior of the critical coupling $\beta_c(m_q)$.  In
Fig.~\ref{fig:criticalcoupling} we plot $\beta_c(m_q)$ defined as the peak
position of the chiral susceptibility $\chi_m$.  Solid lines represent fit of
the data to the form (\ref{eq:zg}), which reasonably go through the
data points.
Results for $z_g$ are listed in the first row of Table~\ref{tab:exponents}.
Other susceptibilities yield results consistent with those from $\chi_m$ well
within the errors.

We observe that the values of $z_g$ do not exhibit clear size
dependence, and are in agreement with the theoretical predictions
based on O(2) or O(4) symmetry within one to two
standard deviations.  As expected from this observation, a reasonable fit
is also obtained fixing $z_g$ to either the O(2) or O(4) exponent.

\begin{figure*}
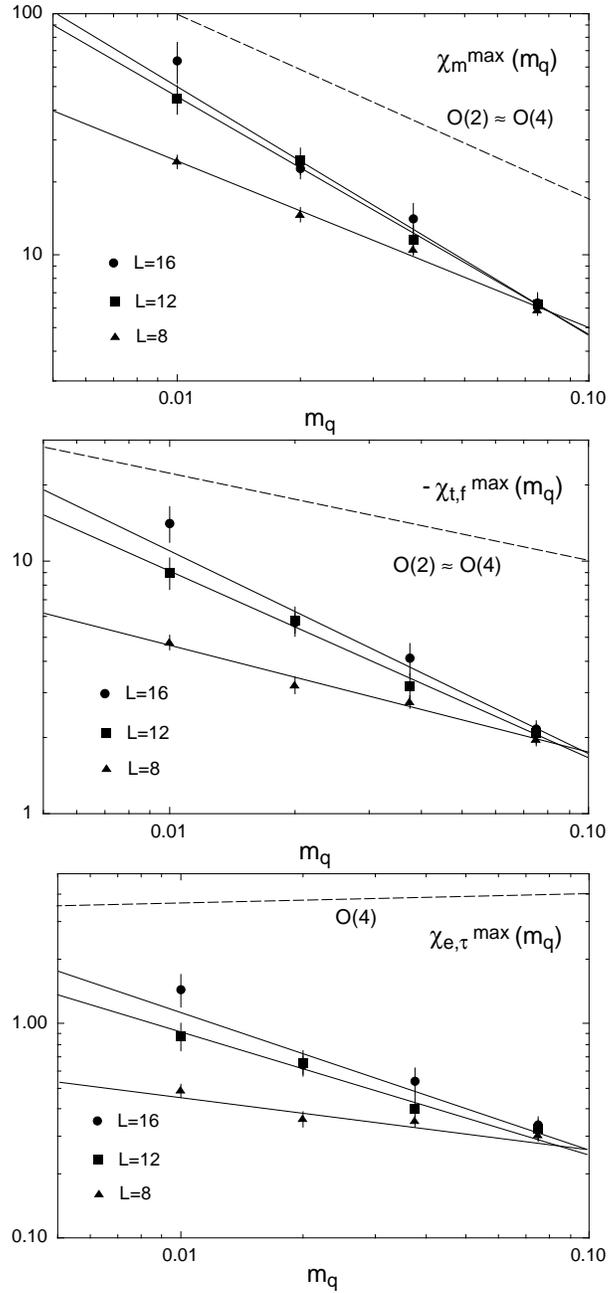

\centerline{\epsfxsize=80mm \epsfbox{peakheight-a.epsf}}
\centerline{\epsfxsize=80mm \epsfbox{peakheight-b.epsf}}
\centerline{\epsfxsize=80mm \epsfbox{peakheight-c.epsf}}
\caption{Peak height of susceptibility $\chi_m$, $\chi_{t,f}$ and
$\chi_{e,\tau}$  as a function of $m_q$ for
fixed spatial size $L$.  Lines are fits to a single power
$\chi_i\propto m^{-z_i}$.  Dashed lines indicate slope expected
for O(2) or O(4) exponents.}
\label{fig:peakheight}
\end{figure*}

Let us turn to the exponents determined from peak height of the
susceptibilities.  In Fig.~\ref{fig:peakheight} we plot the quark mass
dependence of peak height for the representative susceptibilities.
Exponents are
extracted by fits employing a scaling behavior with a single power as given in
(35, 37-39).  Results are summarized in
Table~\ref{tab:exponents}.   For $z_t$ and $z_e$ various operator
combinations yield results which are in mutual agreement within estimated
errors.

\begin{table}[t]
\begin{center}
\caption{Critical exponents extracted by fits of critical coupling and
peak height of susceptibilities for fixed spatial size $L$ as compared to
$O(2), O(4)$\cite{baker,guillou,kanaya} and mean-field (MF) values. }
\label{tab:exponents}
\vspace*{6mm}
\begin{tabular}{lllllll}
\hline
	&O(2)	&O(4)	&MF			&$L=8$	&$L=12$	&$L=16$\\
\hline
$z_g$	&0.60 	&0.54	&2/3			&0.70(11)&0.74(6)&0.64(5)\\
\hline
$z_m$	&0.79	&0.79	&2/3			&0.70(4)&0.99(8)&1.03(9)\\
\hline
$z_t$	&0.39	&0.33	&1/3		\\
$z_{t,f}$		&	&	&	&0.42(5)&0.75(9)&0.78(10)\\
$z_{t,\sigma}$		&	&	&	&0.47(5)&0.81(10) &0.82(12)\\
$z_{t,\tau}$		&	&	&	&0.47(5)&0.81(9) &0.83(12)\\
\hline
$z_e$	&-0.01	&-0.13	&0		\\
$z_{e,f}$		&	&	&	&0.21(4)&0.28(7)&0.38(7)\\
$z_{e,\sigma}$		&	&	&	&0.25(6)&0.56(11) &0.58(13)\\
$z_{e,\tau}$		&	&	&	&0.22(6)&0.52(10) &0.55(12)\\
$z_{e,\sigma\sigma}$	&	&	&	&0.18(5)&0.46(8) &0.43(10)\\
$z_{e,\sigma\tau}$	&	&	&	&0.20(5)&0.51(9) &0.50(12)\\
$z_{e,\tau\tau}$	&	&	&	&0.19(5)&0.48(9) &0.47(11)\\
\hline
\end{tabular}
\end{center}
\end{table}

We observe that all the exponents $z_m, z_t$ and $z_e$ exhibit a sizable
increase between $L=8$ and 12, and the larger values stay for $L=16$.
Comparing the exponents with those of
O(2), O(4) or mean-field (MF) predictions,
we find that an apparent agreement of $z_m$ and $z_t$ for the smallest
size $L=8$ becomes lost for $L\geq 12$.  The magnitude of discrepancy is
smallest for $z_m$, for which we find a 10--20\% larger value amounting to
a one to two standard deviation difference.
For $z_t$ the discrepancy is by a factor two for $L=12$ and 16.
The disagreement is even more pronounced for the exponent $z_e$
for which  a value in the range $z_e\approx
0.5-0.6$ is obtained in contrast to negative values for O(2) and O(4).

One can ask how inclusion of subleading singularities and/or analytic
terms in the fitting function modifies the results above.  A thorough
examination of this question is difficult with our limited data sets, and we
restrict ourselves to the simplest case where a constant term is included in
the fit:
$\chi_i^{max}(m_q)=c_{i0}+c_{i1}m_q^{-z_i}$.
The points to be examined are (i) how the values of exponents change, and
(ii) whether reasonable fits are obtained with the exponents fixed to
theoretically expected values.

Concerning (i), the fitted values of $z_m$ and $z_t$ for $L=8$ and 12
are consistent with the results of single-power fits, while
those for $L=16$ become larger and take a value
$z\approx 1.5\pm 0.5$.  For $z_e$ large values of such a magnitude are
obtained for all three sizes $L=8$, 12 and 16 with similar errors.
Thus adding a constant term does not alleviate the discrepancy.

Turning to (ii), the quality of fit significantly worsens when one
fixes the value
of exponents to the theoretical value.  Values of $\chi^2$ per degree of
freedom increase to $2-3$ as compared to $0.5-1$ for the single-power fit
with $z_i$ as a free parameter, and the fit generally misses the
point for the smallest quark mass $m_q=0.01$ for $L=16$.
In particular the fit for $\chi_e$ accommodates a negative value of $z_e$
only by forcing the coefficient $c_{e,1}$ in front of the power term to a
negative value of a magnitude similar to that of the constant term $c_{e,0}$.
Altogether fits with theoretical values of exponents do not appear any more
reasonable than fits with a single power.

These examinations lead us to conclude that the
exponents do show a trend of deviation from the O(2) or O(4) values,
at least in the range of quark mass $m_q=0.075-0.01$ explored
in our simulation.

\begin{figure*}
\centerline{\epsfxsize=80mm \epsfbox{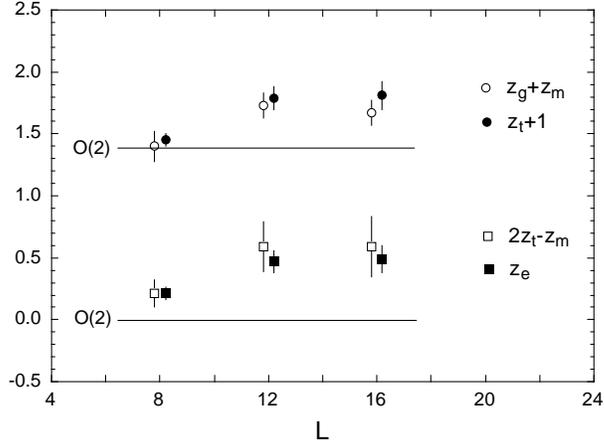}}
\caption{Consistency check of exponents for a given spatial size $L$.
Horizontal lines indicate values expected for O(2) exponents. Values
for O(4) are similar. }
\label{fig:consistency}
\end{figure*}

Let us recall from Sec.~\ref{sec:scaling} that the four exponents
$z_g$, $z_m$, $z_t$ and $z_e$ should satisfy two consistency equations
(\ref{eq:consistencyone}--\ref{eq:consistencytwo}).  In
Fig.~\ref{fig:consistency}  we plot the two sides of the hyperscaling
equations using the
exponents obtained with a single power fit in
Table~\ref{tab:exponents}. For $z_t$ and $z_e$ we take averages over the
channels as the exponents are mutually consistent.
We observe that the hyperscaling relations are
well satisfied for each spatial volume
even though the values of
individual exponents change from volume to volume and
deviate from theoretical expectations. This implies that
our susceptibility data are consistent  with a second-order transition at
$m_q=0$ governed by the magnetic and thermal operators.

Given this result, we may estimate the magnetic and thermal
exponents through a $\chi^2$ fit of the four exponents $z_{g, m, t, e}$
to the form (34, 36, 40, 41).
Using average values of results in Table~\ref{tab:exponents} for $z_t$
and $z_e$, we find $(y_h, y_t)=(2.31(7),
1.74(5)) (L=8)$, $(3.02(19), 2.24(12))(L=12)$ and
$(3.31(25), 2.22(15))(L=16)$, as compared to
$(2.48, 1.49)$ for
O(2) symmetry and $(2.49, 1.34)$ for O(4) symmetry\cite{baker,guillou,kanaya}.

\subsection{Results for scaling function}

Defining a scaling variable
\be
x=\left(\beta_c(m_q)-\beta_c(0)\right)\cdot m_q^{-z_g},
\ee
one expects the singular part of the susceptibility  to take the
functional form
\be
\chi_m(\beta,m_q)=m_q^{-z_m}\cdot F_m(x).
\ee

\begin{figure*}
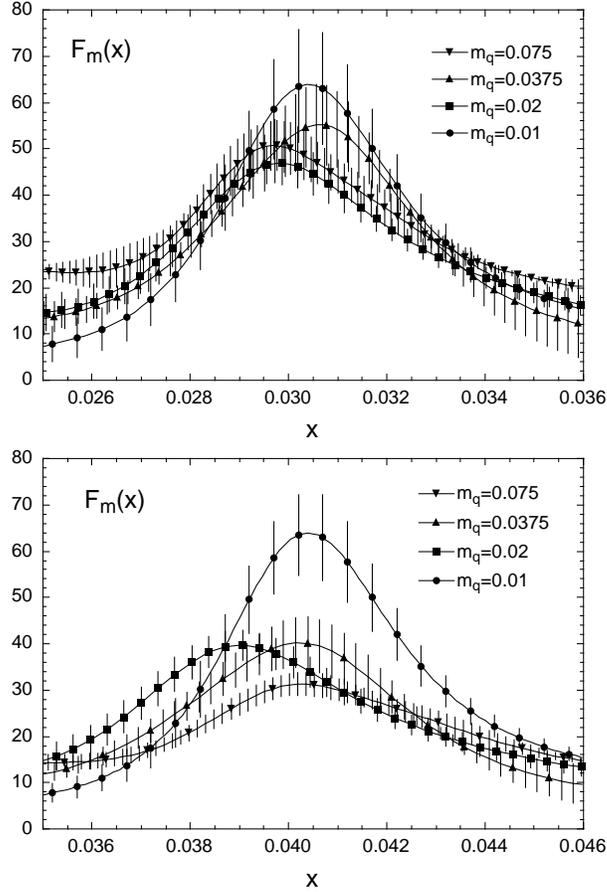

\centerline{\epsfxsize=80mm \epsfbox{scalingfunction-a.epsf}}
\centerline{\epsfxsize=80mm \epsfbox{scalingfunction-b.epsf}}
\caption{Scaling function $F_m(x)$ calculated by
$\chi_m(\beta,m_q)\cdot (m_q/0.01)^{z_m}$ as a function of
$x=\left(\beta_c(m_q)-\beta_c(0)\right)\cdot (m_q/0.01)^{-z_g}$for $L=16$ using
(a) measured
values $z_g=0.6447, z_m=1.033, \beta_c(0)=5.2353$, and (b) O(4) values
$z_g=0.538$ and $z_m=0.794$ and measured value $\beta_c(0)=5.2253$.}
\label{fig:scalingfunction}
\end{figure*}

We plot in Fig.~\ref{fig:scalingfunction} two estimates of the scaling function
$F_m(x)$ calculated as $\chi_m(\beta,m_q)\cdot (m_q/0.01)^{z_m}$ using data for
$L=16$: in (a) we employ the measured values $z_g=0.6447$, $z_m=1.033$,
$\beta_c(0)=5.2353$, and in (b) we take the O(4) values\cite{kanaya}
for the exponents
$z_g=0.538$, $z_m=0.794$ and substitute the value $\beta_c(0)=5.2253$
obtained with a fit of $\beta_c(m_q)$ with $z_g$ fixed to the O(4) value.
Similar
to the experience with fits of peak height in Sec.~\ref{sec:exponents}, we find
that scaling is reasonable with the use of the measured exponents. The fit,
however, worsens if
the O(4) exponents are employed;  in particular the curve for the smallest
quark mass $m_q=0.01$ deviates largely from the rest.

\section{Conclusions}

In this article we have presented results of our analysis of the
two-flavor chiral phase transition with the Kogut-Susskind quark action on an
$N_t=4$ lattice.  By studying the spatial
volume dependence of various susceptibilities, we have confirmed the conclusion
of previous investigations\cite{founf2,columbianf2}
that the transition is a smooth
crossover for $m_q\geq 0.02$.  At $m_q=0.01$ the susceptibilities exhibit an
almost linear increase in spatial volume between $8^3$ and $16^3$ lattices,
which contradicts the results of previous work\cite{columbianf2}, and may
appear to be
consistent with a first-order phase transition.  However, examination of
time histories and histograms of observables, and in particular,
a rescaling of spatial size in
terms of the zero-temperature pion mass strongly suggests that
the linear increase is a transient phenomenon arising from
an insufficient spatial size.
It is our present conclusion that there is no evidence
indicating a first order transition  down to $m_q=0.01$.


We have also analyzed how susceptibilities depend on the quark mass.  The
pattern of critical exponents we have obtained is consistent with the existence
of a second-order phase transition at $m_q=0$, which is governed by a
renormalization-group fixed point with two relevant operators,
the energy and magnetization operators.
The exponents, however, do not agree
with  O(2), O(4) or mean-field theory predictions.
This means that the theoretical argument for a second order phase
transition from the chiral sigma model remains unjustified in the
present work.

A disagreement with the O(4) values may not come as a surprise since flavor
symmetry breaking effects of the Kogut-Susskind quark action is quite large at
$\beta\approx 5.3$ where the transition is located for $N_t=4$.  Indeed masses
of non-Nambu-Goldstone pions are closer to those of $\rho$ meson, rather
than those of the Nambu-Goldstone pion, for these
values of $\beta$.

Numerically, the O(2) values for exponents are close to those for O(4).
The deviation from the O(2) values is theoretically more puzzling for
several reasons:
(i) O(2) is an exact symmetry group of the Kogut-Susskind action
for any lattice spacing,  (ii) this symmetry is preserved under the algorithmic
expedient of taking a square root of the quark determinant adopted in the
hybrid
R algorithm, and (iii) the susceptibility $\chi_m$ is precisely the second
derivative of free energy with respect to the quark mass which is the conjugate
field of the O(2) order parameter.  Thus, if the two-flavor system
simulated by the hybrid R algorithm undergoes a second-order transition,
we expect the O(2) values of exponents to emerge toward the chiral limit.

The smallest quark mass $m_q=0.01$ we have explored is quite small
at $\beta\approx 5.3$,
corresponding to $m_\pi/m_\rho\approx 0.19$ which is close to the experimental
value of 0.18.  It is possible, however, that the critical region where
susceptibilities exhibit the true scaling behavior is located even nearer
to the chiral limit.  If this is
the origin of the discrepancy, establishing the universality nature of the
two-flavor transition for the Kogut-Susskind quark action will require further
simulations toward substantially smaller quark masses and necessarily
much larger
spatial lattices.

\section*{Acknowledgements}

We would like to thank Frithjof Karsch and Edwin Laermann
for informative discussions.
This work is supported by the Supercomputer Project (No. 97-15)
of High Energy Accelerator Research Organization (KEK),
and also in part by the Grants-in-Aid of
the Ministry of Education (Nos. 08640349, 08640350, 08640404,
09246206, 09304029, 09740226).

\end{document}